\documentclass[12pt]{iopart}

\newcommand{\cs}{\left| \alpha\rangle\langle\alpha \right|}
\newcommand{\al}{\alpha}

\newcommand{\pa}{\partial}

\usepackage{amssymb}
\usepackage{iopams}
\usepackage{footnote}

\begin{document}

\title[Yu.\,E.\,Kuzovlev\,:\, %
Electron diffusivity 1/f noise]{Electron %
random walk in ideal phonon gas. \\
Spectra of density matrix evolution \\ %
and electron mobility 1/f\, noise}

\author{Yu. E. Kuzovlev}

\address{Donetsk Institute
for Physics and Technology of NASU, 83114 Donetsk, Ukraine}
\ead{kuzovlev@fti.dn.ua}

\begin{abstract}
The previously derived exact evolution equations for density %
matrix of electron (quantum particle) in phonon field (boson %
thermostat) are qualitatively analysed. Their statistical %
interpretation is explained in detail, and their main symmetry %
and spectral properties are expounded. In application to the %
electron's random walk, it is shown that these properties %
certaimly forbid conventionally assumed Gaussian long-range %
asymptotic of the walk statistics. Instead, the exact equations %
imply super-linear dependence of fourth-order cumulant of total %
electron's path on observation time, which signifies existence of %
1/f-type low-frequency fluctuations in electro's duffusivity and %
mobility. Physical meaning of this result is discussed, along with %
general origin of 1/f-noise in classical and quantum Hamiltonian %
many-particle systems.
\end{abstract}


\vspace{2pc} \noindent{\it Keywords}\,: %
Dynamical foundations of kinetics, %
Quantum kinetic equations, %
Electron-phonon interaction, %
Electron mobility fluctuations, 1/f noise

\section{Introduction}

This paper  is direct continuation %
(second part) of the work \cite{july}, %
where exact equation for a quantum particle (``electron'') %
in interaction with harmonic boson thermostat %
(``ideal phonon gas'') were reduced to a visual %
shortened form lightening their consideration, %
and, besides, their relation to 1/f noise (1/f\,- type %
fluctuations of electron's mobility) was discussed. %

Here, we start systematic investigation of these %
equations. First, we will consider more carefully their %
statistical meaning, as it looks from the viewpoint %
of probability theory. Then we will formulate their %
main principal symmetry properties and spectral %
properties and show that they are incompatible with %
such ``white-noise'' statistics of the electron's %
random walk what follows from standard kinetic %
approximations. %
The alternative predicted by the exact equations %
is such statistics which %
combines white noise in velocity of electron and %
1/f noise (or ``flicker'' noise) in its diffusivity %
and mobility.

We will finish the paper by discussion %
of this our concrete result in the framework of one more %
attempt to make our understanding of universal nature %
of 1/f noise being clear also for scientific %
community.

\section{The model and the problem}

First, let us recollect main results of \cite{july}. %
We have considered the simplest Hamiltonian of %
quantum particle (``electron'') in quantum boson %
(``phonon'') field,
\begin{eqnarray}
H\,=\,H_e+H_{ph} +H_{int}\,\,\,,\nonumber\\
H_e=\frac {p^2}{2m}\,\,\,, \,\,\,\,\,\, %
H_{ph}=\sum_k\, \hbar\omega_k\, a_k^\dagger a_k\,\,\,,\nonumber\\
H_{int}= \frac 1{\sqrt{\Omega}} \sum_k\, %
[\,c_k^* \,e^{\,ikr}a_k\,+\,c_k\,e^{-ikr} a_k^\dagger \,] %
\,\,\,, \label{h}
\end{eqnarray}
and under the thermodynamical limit, when the system's %
volume $\,\Omega\rightarrow\infty\,$, %
derived the following exact %
hierarchy of evolution equations:
\begin{eqnarray}
\dot{\Delta}_n \, =\,  %
-\,\widehat{V} \nabla_X\, \Delta_n\, %
-\,i\sum_{k\,\in\,K_n}\,  \sigma_k\, %
(\omega_k\,-\,k\widehat{V})\, \Delta_n %
\,-\,\nonumber\\ %
-\,(i\hbar\sqrt{\Omega_0})^{-1} %
\sum_{k\,\in\,K_n}\, %
\sigma_k\, c_k^{\sigma_k}\, %
[\,e^{\,ikY/2}-e^{-ikY/2}(1+N^{-1}_k)\,]\, %
\Delta_{n-1}\,+ \label{edn}\\ %
+ \,\frac {\sqrt{\Omega_0}}{i\hbar}\,  %
\sum_{\sigma\in \{+,-\}}\, \sigma %
\int c_q^{-\sigma}\, %
(\,e^{\,iqY/2}-e^{-iqY/2})\, N_q \,\,  %
\Delta_{n+1}\, \,\frac {d^3q}{(2\pi)^3}\, %
\,\, \nonumber
\end{eqnarray}
Here

$\,\Delta_0=\Delta_0(t,X,Y)\,$ is %
density matrix of the electron, %
in the coordinate representation, %
and $\,\Delta_n=\Delta_n(t,X,Y,K_n,\Sigma_n)\,$ %
are definite functions characterizing statistical %
correlations of electron with  %
quasi-classic (coherent-state) %
amplitudes $\,A_k^\sigma\,$ %
($\,A_k^-\equiv A_k\equiv a_k\,\exp{(ikX)}\,$, %
, $\,A_k^+\equiv A_k^* \equiv a_k^*\,\exp{(-ikX)}\,$) %
of  $\,n\,$ phonon modes marked by %
$\,K_n=\{k_1\dots k_n\}\,$\, and\, %
and\, $\,\Sigma_n =\{\sigma_1\dots\sigma_n\}\,$\,, %
with\, $\,\sigma_j=\pm\,$\, (or\, $\,\pm1\,$);\, %

$\,K_{n-1}=K_n\ominus k\,$ and %
$\,\Sigma_{\,n-1}=\Sigma_{\,n}\ominus \sigma_k\,$ %
in the third term on the right, and %
$\,K_{n+1}=K_n\oplus q\,$ and %
$\,\Sigma_{\,n+1}=\Sigma_{\,n}\oplus \sigma\,$ %
in the last tern (``collision integral'');\, %

$\,X\,$ is electron's coordinate, and $\,Y\,$ is %
spatial variable conjugated with %
electron's momentum $\,p\,$\,, %
so that $\,\widehat{V}=-i(\hbar/m)\nabla_Y\,$ is electron's %
velocity operator, and %
the Fourier transform\, %
\[
\int \exp{(-ipY/\hbar)}\, \Delta_n\, %
d^3 Y/(2\pi\hbar)^3\, %
\]
leads to the Wigner representation;\,

$\,N_k=[\exp{(\hbar \omega_k/T)}-1]^{-1}\,$ %
are equilibrium occupancies  of phonon modes;\, %

$\,c^-_k=c^*_k,\,c^+_k=c_k\,$,\,

and $\,\Omega_0\,$ is %
some fixed (formally arbitrary) volume.\, %

Initial conditions to Eqs.\ref{edn} can be taken %
in the form
\begin{eqnarray}
\Delta_n(t=0)\,=\, \delta_{n\,0}\, %
\Delta_0^0\,\,\, \label{icd}
\end{eqnarray}
(one particular reasonable choice of %
$\,\Delta_0^0=\Delta_0^0(X,Y)\,$ %
was mentioned in \cite{july}), or  %
\begin{eqnarray}
\Delta_n(t=0)\,=\, W_0(X)\, %
\Delta_n^{eq}(Y,K_n,\Sigma_n)\,\,\, \label{icd1}
\end{eqnarray}
with $\,\Delta_n^{eq}\,$ being stationary %
(equilibrium and spatially homogeneous) solution %
of Eqs.\ref{edn} (normalized to %
$\,\Delta_0^{eq}(Y=0)=1\,$).

Foe more details, please, see \cite{july}.%

Notice that derivation of Eqs.\ref{edn} there %
does not touch the purely electron's part of full quantum %
Liouvile (von Neumann) operator, %
therefore the resulting equations %
can be trivially generalized %
to electron under action of an external force $\,f\,$ %
(switched on at $\,t=0\,$) by mere adding terms %
$\,(-fY/i\hbar)\,\Delta_n\,$ to right-hand %
sides of (\ref{edn}).

Our main problem is qualitatively complete and %
quantitatively correct analysis of large-time evolution %
of electron's position (and total path) probability %
distribution, %
\[
W(t,X)\,=\,\Delta_0(t,X,Y=0)\,\,\,,
\]
if initially the electron was located in some finite %
space region, e.g. in vicinity of $\,X=0\,$.

\section{Statistical contents of the evolution %
equations}

To perceive more carefully statistical meaning of %
the functions $\,\Delta_n\,$ ($\,n>0\,$),  %
let us return to their basic definition, - %
i.e. formula (21) from \cite{july}, - %
and rewrite it in equivalent form
\begin{eqnarray}
\Delta_n\,=\, \lim_{\Omega\,\rightarrow\,\infty}\,%
\int F(A)\, \left[ %
\prod_{k\,\in\,K_n}\, %
\frac {\sqrt{\Omega}}{\sqrt{\Omega_0}} \cdot  %
\frac {\pa }{\pa A_{k}^{\,\sigma_k}}\,\,  %
\right]\, F^{-1}(A)\,P(A)\,dA %
\,\,\,, \label{d}
\end{eqnarray}
where $\,P(A)=P(t,X,Y,A)\,$ %
is the quasi-probability %
density distribution of all the phonon %
amplitudes $\,A\equiv\{A_k^-,A_k^+\}\,$,\, %
$\,\int\dots dA\,$ is integral over all them, %
and
\[
F(A)\,\equiv\,\exp{\left(-\sum_k\, %
\frac {A_k^- A_k^+}{N_k}\right)}\,\,  %
\]
The equivalence directly follows from %
the expansion (20) %
in \cite{july} (along with relations between %
functions $\,P,\,Q,\,P_n\,$ and $\,Q_n\,$ %
pointed out there). %
The integration by parts turns (\ref{d}) into %
\begin{eqnarray}
\Delta_n\,=\, \lim_{\Omega\,\rightarrow\,\infty}\,%
\int P(A) \,\left\{\, F^{-1}(A)\,  %
\prod_{k\,\in\,K_n} \left[ %
-\frac {\sqrt{\Omega}}{\sqrt{\Omega_0}} \cdot  %
\frac {\pa }{\pa A_{k}^{\,\sigma_k}}\,\,  %
\right]\, F(A)\,\right\}\, dA %
\,\,\, \label{d1}
\end{eqnarray}
If all wave vectors of the set $\,K_n\,$ %
are different one from another then this expression %
reduces to formula (38) from \cite{july}. %
We now will consider general case allowing %
coincidence of some of the wave vectors %
even in the thermodynamic limit, that is after %
transition from discrete to continuous %
phonon spectrum.  %

Symbolically,  this transition reduces to %
mere re-scaling of the phonon amplitudes, %
so that
\begin{eqnarray}
\sqrt{\Omega}\,A^\sigma_k\,\Rightarrow\, A^\sigma_k\,\,, %
\,\,\,\,\,\, %
\sqrt{\Omega}\, \frac \pa{\pa A^\sigma_k}\,\Rightarrow\, %
(2\pi)^3\,\frac \delta{\delta A^\sigma_k}\,\,, %
\nonumber\\
\frac 1\Omega\sum_k\,\dots \,\Rightarrow\, %
\int \dots\, \frac {d^3k}{(2\pi)^3}\,\,\,, \nonumber\\
F(A)\,\Rightarrow\,\exp{\left(-\int %
\frac {A_k^- A_k^+}{N_k}\, %
\frac {d^3k}{(2\pi)^3}\right)}\,\, \,, %
\label{f}
\end{eqnarray}
and similar redefinition of the parent creation %
and annihilation operators, %
so that their commutator turns to %
\[
[a_k,a_q^\dagger ]\,\Rightarrow\, %
(2\pi)^3\,\delta(k-q)\,\,\, %
\]
(see e.g. \cite{july}) %
while phonon-related part the Hamiltonian (\ref{h}) to %
\begin{eqnarray}
H_{ph}=\int \hbar\omega_k\, a_k^\dagger a_k\,dk\,\,, %
\,\,\,\,\, %
H_{int}= \int %
[\,c_k^* \,e^{\,ikr}a_k\,+\,c_k\,e^{-ikr} a_k^\dagger \,] %
\,dk \,\,\,, \label{h1}
\end{eqnarray}
with\, $\,dk\,\equiv\,d^3k/(2\pi)^3\,$\,.\, %
Consequently, instead of (\ref{d1}) we can write %
\begin{eqnarray}
\Delta_n\,=\,
\int P(A) \,\left\{\, F^{-1}(A)\,  %
\prod_{k\,\in\,K_n} \left[ %
-\frac {(2\pi)^3}{\sqrt{\Omega_0}} \cdot  %
\frac {\delta }{\delta A_{k}^{\,\sigma_k}}\,\,  %
\right]\, F(A)\,\right\}\, dA %
\,\,\,, \label{d2}
\end{eqnarray}
where now $\,\int P(A)\,\{\dots\}\,dA\,$ means %
functional integration.

The Eqs.\ref{d1} and \ref{d2}, together with %
the mentioned formula (38) from \cite{july}, %
prompt to redefine functions $\,\Delta_n\,$ %
as follows:
\[
\Delta_n(t,X,Y,K_n,\Sigma_n)\,\Rightarrow\, %
\Delta_n(t,X,Y,K_n,-\Sigma_n)\, %
\prod_{k\,\in\,K_n}\, %
\frac 1{N_k\sqrt{\Omega_0}}\, %
\]
Then  our basic Eqs.\ref{edn} take form
\begin{eqnarray}
\dot{\Delta}_n \, =\,  %
-\,\widehat{V} \nabla_X\, \Delta_n\, %
+\,i\sum_{k\,\in\,K_n}\,  \sigma_k\, %
(\omega_k\,-\,k\widehat{V})\, \Delta_n %
\,+\,\nonumber\\ %
+\,(i\hbar)^{-1} %
\sum_{k\,\in\,K_n}\, %
\sigma_k\, c_k^{-\sigma_k}\, %
[\,N_k\,e^{\,ikY/2}- (N_k +1)\,e^{-ikY/2}\,]\, %
\Delta_{n-1}\,- \label{ne}\\ %
- \,(i\hbar)^{-1}\,  %
\sum_{\sigma\in \{+,-\}}\, \sigma %
\int c_q^{\,\sigma}\, %
(\,e^{\,iqY/2}-e^{-iqY/2})\,\,  %
\Delta_{n+1}\, \,\frac {d^3q}{(2\pi)^3}\,\,\,, %
\,\, \nonumber
\end{eqnarray}
where, as above,\, $\,\Delta_n=\Delta_n(t,X,Y,K_n, %
\Sigma_n)\,$,\, %
$\,\Delta_{n-1}=\Delta_{n-1}(t,X,Y,K_n\ominus k, %
\Sigma_n\ominus \sigma_k)\,$,\, %
$\,\Delta_{n+1}=\Delta_{n+1}(t,X,Y,K_n\oplus q, %
\Sigma_n\oplus \sigma)\,$\,.  %
 We may write them in compact form
\begin{eqnarray}
\dot{\Delta}\,=\, - %
\widehat{V}\nabla_X\, \Delta\,+\, %
\widehat{\Lambda}\, \Delta\,\,\,, \label{nes}
\end{eqnarray}
with $\,\Delta=\{\Delta_0,\Delta_1,\dots\}\,$ %
and operator\,  $\,\widehat{\Lambda}= %
\widehat{\Lambda}(Y,\nabla_Y)\,$\, %
unifying all the phonon-related operators %
from right-hand sides of (\ref{ne}). %
Then Eq.\ref{d2} yields
\begin{eqnarray}
\Delta_n\,=\,
\int P(A) \,\left\{\, F^{-1}(A)\,  %
\prod_{k\,\in\,K_n} \left[ %
-\,(2\pi)^3\,N_k\,  %
\frac {\delta }{\delta A_{k}^{-\sigma_k}}\,\,  %
\right]\, F(A)\,\right\}\, dA %
\,\,\, \label{nd}
\end{eqnarray}
In the case of non-coinciding wave vectors %
this reduces to
\begin{eqnarray}
\Delta_n\,=\, \int P(A) \,\left\{\, %
\prod_{k\,\in\,K_n}\, A_{k}^{\,\sigma_k}\, %
\right\}\, dA\,\,\,, \label{dm}
\end{eqnarray}
or, equivalently,
\begin{eqnarray}
\Delta_n\,=\, \left\langle\, %
\delta(X(t)-X)\,\exp{(iYp(t)/\hbar)}\,
\prod_{k\,\in\,K_n}\, A_{k}^{\,\sigma_k}(t)\, %
\right\rangle\,\,\, \label{dmr}
\end{eqnarray}
in the coordinate representation, or
\begin{eqnarray}
\Delta_n\,=\, \left\langle\, %
\delta(X(t)-X)\,\delta(p(t)-p)\,
\prod_{k\,\in\,K_n}\, A_{k}^{\,\sigma_k}(t)\, %
\right\rangle\,\,\, \label{dmp}
\end{eqnarray}
in the Wigner representation, %
where\, $\,X(t)\,$, $\,p(t)\,$ and %
$\,A_{k}^{\,\sigma}(t)\,$\, mean electron's coordinate %
and momentum and phonon amplitudes considered %
as random processes.

If some of wave vectors from $\,K_n\,$ are equal %
and, besides, their  counterparts $\,\sigma_k=\pm\,$ %
have opposite signs, then - in contrast %
to (\ref{dm}), - the braces in (\ref{d1}),
(\ref{d2}) and (\ref{nd}) %
represent definite Hermite polynomials %
of phonon amplitudes. Statistical meaning %
of the resulting expressions becomes clear %
if one unifies all the redefined %
functions  $\,\Delta_n\,$'s, - %
with arbitrary $\,K_n\,$'s, - into %
generating functional
\begin{eqnarray}
\Delta\,\equiv\, \Delta_0\,+\sum_{n\,=\,1} %
^\infty \,\frac 1{n!} %
\sum_{\sigma_1\dots \sigma_n}\,\int %
\Delta_n\, \prod_{j\,=\,1}^n\, %
z_{\sigma_j}(k_j)\,d^3k_j\, %
\,\, \label{gf}
\end{eqnarray}
Application of Eq.\ref{nd} gives
\begin{eqnarray}
\Delta\,=\, %
\exp{\left[-\int (2\pi)^3N_k\, %
z_+(k)z_-(k)\,d^3k\,\right]}\, %
\times\,\nonumber\\%
\times\, \int \exp{\left[\sum_\sigma \int %
z_\sigma(k) A^\sigma_k \,d^3k\,\right]}\,
P(A)\,dA \,\, \label{fm}
\end{eqnarray}
First exponential on the right here is nothing %
but inverse of equilibrium characteristic %
functional of free phonon field %
(i.e. in absence of the electron:
\begin{eqnarray}
\exp{\left[\int (2\pi)^3N_k\,\, %
z_+(k)z_-(k)\,d^3k\,\right]}\,=\,\nonumber\\ %
=\,\int \exp{\left[\sum_\sigma \int %
z_\sigma(k) A^\sigma_k \,d^3k\,\right]}\,
P_{free}(A)\,dA\,\equiv\, \, \label{fm0}\\
\equiv\, %
\left\langle\, %
\exp{\left[\sum_\sigma \int %
z_\sigma(k) A^\sigma_k \,d^3k\,\right]}\,
\right\rangle_{free}\, %
\equiv\, \mathcal{F}_{free}\{z\} \,\,\,,\nonumber
\end{eqnarray}
where
\[
P_{free}(A)\,=\,F(A)\,\left[\int F(A)\,dA %
\right]^{-1}\,
\]
is continuous limit of the discrete measure
\[
\prod_k\,f_k\,=\,\prod_k\,%
(2\pi N_k)^{-1}\exp{(-|A_k|^2/N_k)}\,
\]
(see \cite{july}). %
Hence, in the frames of coordinate representation %
Eq.\ref{fm} can be written as
\begin{eqnarray}
\Delta\,=\, \Delta\{t,X,Y,z\}\, %
=\, \frac {\mathcal{F}\{t,X,Y,z\}} %
{\mathcal{F}_{free}\{z\}}\,\,\,,\label{fmr}\\
\mathcal{F}\{t,X,Y,z\}\,=\,\label{ffr}\\ %
=\,\left\langle\, %
\delta(X(t)-X)\,%
\exp{\left[\frac {iYp(t)}\hbar \right]}\,
\exp{\left[\sum_\sigma \int %
z_\sigma(k) A^\sigma_k(t) \,d^3k\,\right]} %
\,\right\rangle\, \nonumber
\end{eqnarray}
And in the Wigner representation
\begin{eqnarray}
\Delta\,=\, \Delta\{t,X,p,z\}\, %
=\, \frac {\mathcal{F}\{t,X,p,z\}} %
{\mathcal{F}_{free}\{z\}}\,\,\,,\label{fmp}\\
\mathcal{F}\{t,X,p,z\}\,=\,\label{ffp}\\ %
=\, \left\langle\, %
\delta(X(t)-X)\,\delta(p(t)-p)\,
\exp{\left[\sum_\sigma \int %
z_\sigma(k) A^\sigma_k(t) \,d^3k\,\right]} %
\,\right\rangle\, \nonumber
\end{eqnarray}
Formulae (\ref{fm})-(\ref{fmp}) give %
complete and transparent statistical interpretation of %
generating functional of $\,\Delta_n\,$'s %
and thus $\,\Delta_n\,$'s themselves. %
It remains to notice that in (symbolic) terms %
of the full density matrix of the system, %
$\,\rho\,$ (see formula (4) in [1]), %
\begin{eqnarray}
\mathcal{F}\{t,X,Y,z\}\,=\, %
\Tr_{\,ph}\, %
\exp{\left[\int %
z_-(k) \,a_k\,e^{ikr} \,d^3k\,\right]}\, %
\times\nonumber\\
\times \,\langle r|\,\rho\,|r^\prime\rangle \, %
\exp{\left[\int %
z_+(k) \,a_k^\dagger\, %
e^{-ikr^\prime} \,d^3k\,\right]} %
\,\,\,, \label{fmo}
\end{eqnarray}
where\, $\,r=X+Y/2\,$,\, %
$\,r^\prime = X-Y/2\,$,\, %
$\,|r\rangle\,$ and $\,|r^\prime\rangle\,$ %
are eigenstates of the electron's coordinate %
operator, and $\,\Tr_{\,ph}\,$ %
is trace over all phonon states.

\section{Generating evolution equation}

The Eqs.\ref{ne} are useful equivalent of %
Eqs.37 in [1] for functions $\,D_n\,$ %
representing the same correlations  %
as $\,\Delta_n\,$ in Eqs.\ref{dmr}  and \ref{dmp} %
but in dimensionless relative units %
(obviously,
\[
\Delta_n(K_n,\Sigma_n)\,=\,D_n(K_n,-\Sigma_n)\, %
\prod_{k\,\in\,K_n}\, \left( %
-\frac {c_k^{-\sigma} }{\hbar\omega_k}\right)\,\,\,, %
\]
after our above redefinition of $\,\Delta_n\,$). %
Corresponding equivalent of the %
functional evolution equation %
(40) from [1] is %
\begin{eqnarray}
\dot{\Delta} \, =\,  %
-\,\widehat{V} \nabla_X\, \Delta\, %
+\,\sum_\sigma \int d^3k\,\,\, z_\sigma(k)\, %
\widehat{L}_{k\,\sigma}\,\frac %
\delta{\delta z_\sigma(k)}\,\,\Delta\, %
+\, \label{fe}\\
+\, \sum_\sigma \int d^3k\,\,\, z_\sigma(k)\, %
\widehat{B}_{k\,\sigma}\, %
\,\Delta\,+\, %
\sum_\sigma \int d^3k\,\,\, %
\widehat{A}_{k\,\sigma}\,\frac %
\delta{\delta z_\sigma(k)}\,\,\Delta\, %
\,\equiv\, \nonumber\\
\equiv\, %
-\,\widehat{V} \nabla_X\, \Delta\, %
+\, \widehat{\mathcal{L}} %
\left\{z,\frac \delta{\delta z} %
\right\}\,\Delta\, %
\,\,, \nonumber
\end{eqnarray}
where now
\begin{eqnarray}
\widehat{L}_{k\,\sigma}\,\equiv\, %
\,i  \sigma\, %
(\omega_k\,-\,k\widehat{V})\,\,\,, \nonumber\\
\widehat{B}_{k\,\sigma}\,\equiv\, %
\frac {\sigma\,c_k^{-\sigma} }{i\hbar} \, %
[\,e^{\,ikY/2}N_k\,-\,e^{-ikY/2}(N_k+1)\,]\, %
\,\,, \label{fops} \\
\widehat{A}_{k\,\sigma}\,\equiv\, %
- \frac {c_k^{\,\sigma}}{i\hbar\, (2\pi)^3}\, %
\,(\,e^{\,ikY/2}-e^{-ikY/2})\,  %
\,\, . \nonumber
\end{eqnarray}

Correspondingly, the functional $\,\mathcal{F}\,$ %
undergoes equation %
\begin{eqnarray}
\dot{\mathcal{F}}\,=\, %
-\,\widehat{V} \nabla_X\, \mathcal{F}\, %
+\, \widehat{\mathcal{L}}_0 %
\left\{z,\frac \delta{\delta z} %
\right\}\,\mathcal{F}\,\,\,, \label{fe0} %
\end{eqnarray}
where
\begin{eqnarray}
\widehat{\mathcal{L}}_0\,\equiv\, %
\mathcal{F}_{free}\, %
\widehat{\mathcal{L}}\,\mathcal{F}_{free}^{-1}\, %
=\, \nonumber\\ %
=\,\sum_{sigma} \int d^3k\,\,\, z_\sigma(k)\, %
\widehat{L}_{k\,\sigma}\,\frac %
\delta{\delta z_\sigma(k)}\,\,\, %
+\, \label{fop0}\\
+\, \sum_\sigma \int d^3k\,\,\, z_\sigma(k)\, %
\widehat{B}^0_{k\,\sigma}\, %
\,+\, %
\sum_\sigma \int d^3k\,\,\, %
\widehat{A}_{k\,\sigma}\,\frac %
\delta{\delta z_\sigma(k)}\, %
\,\,, \nonumber
\end{eqnarray}
with
\begin{eqnarray}
\widehat{B}^0_{k\,\sigma}\,\equiv\, %
-\frac {\sigma\,c_k^{-\sigma}}{i\hbar} \, %
\,e^{-ikY/2}\, %
\,\, \label{b0}
\end{eqnarray}
Notice that the evolution operator %
$\,\widehat{\mathcal{L}}_0\,$ does not %
include the phonon mode occupancies %
(which instead must appear in initial %
conditions for $\,\mathcal{F}\,$).

\section{Symmetry properties of %
correlation functions and their evolutions}

According to previous section, both the  hierarchy %
$\,\Delta =\{\Delta_0,\Delta_1,\dots\}\,$ %
as the whole and the functional (\ref{gf}) %
are equivalent to full density matrix of  %
our system. The hermicity of this density matrix %
means that %
\begin{eqnarray}
\Delta_n(-Y,-\Sigma_n)\, %
=\, \Delta_n^*(Y,\Sigma_n) \,\,\,, \label{hr}
\end{eqnarray}
in the coordinate representation, or %
\begin{eqnarray}
\widehat{\mathcal{H}}\,\Delta_n(Y,\Sigma_n)\, %
\equiv\, \Delta_n^*(-Y,-\Sigma_n)\, %
=\, \Delta_n(Y,\Sigma_n) \,\, \nonumber
\end{eqnarray}
Here and below we write out only %
arguments what are under current attention. %
In terms of the generating functional, %
\begin{eqnarray}
\widehat{\mathcal{H}}\, \Delta\{Y,z_{\sigma}\}\, %
\equiv\, \Delta^*\{-Y,z_{-\sigma}\}\,=\,
\Delta\{Y,z_{\sigma}\}\,\,\,, \label{hrf}
\end{eqnarray}
where\, $\,*\,$\, means complex conjugation of %
coefficients of the functional %
(but not its argument $\,z_{\sigma}(k)\,$). %
Indeed, one can verify that evolution operator %
in Eqs.\ref{ne} and \ref{fe} satisfies %
\begin{eqnarray}
\widehat{\mathcal{H}}\, %
[\,-\widehat{V}\nabla_X\,+\,\widehat{\Lambda}\,]\, %
\widehat{\mathcal{H}}^{-1}\,=\, %
[\,-\widehat{V}\nabla_X\,+\,\widehat{\Lambda}\,] %
\,\,\,,\nonumber\\ %
\widehat{\mathcal{H}}\, %
[\,-\widehat{V}\nabla_X\,+\,\widehat{\mathcal{L}}\,]\, %
\widehat{\mathcal{H}}^{-1}\, %
=\,[\,-\widehat{V}\nabla_X\,+\,\widehat{\mathcal{L}}\,] %
\,\label{ho}
\end{eqnarray}
Hence, if the hermicity property is satisfied at %
any one time moment (e.g. when initial conditions %
(\ref{icd}) take place) then it keeps at all %
other time moments. %
In the Wigner representation the same property looks as
\begin{eqnarray}
\widehat{\mathcal{H}}\,\Delta_n(p,\Sigma_n)\, %
\equiv\, \Delta_n^*(p,-\Sigma_n)\, %
=\, \Delta_n(p,\Sigma_n) \,\,\,, \label{hp}
\end{eqnarray}
with similar changes in (\ref{hrf}) %
and (\ref{ho})\, and with \, %
$\,\widehat{V}\Rightarrow \,p/m\,$\,.

Further, if the system's Hamiltonian %
expressed by (\ref{h}) and (\ref{h1}) %
is invariant in respect to time inversion, - %
which is the case when
\begin{eqnarray}
\omega_{-k}=\omega_k\,\,\,, \,\,\,\,\, c_{-k}=c_k^*\,\,\,,
\label{hrev}
\end{eqnarray}
- then
\begin{eqnarray}
\widehat{\Theta}\, %
[\,-\widehat{V}\nabla_X\,+\,\widehat{\Lambda}\,]\, %
\widehat{\Theta}^{-1}\,=\,-\, %
[\,-\widehat{V}\nabla_X\,+\,\widehat{\Lambda}\,] %
\,\,\,,\nonumber\\ %
\widehat{\Theta}\, %
[\,-\widehat{V}\nabla_X\,+\,\widehat{\mathcal{L}}\,]\, %
\widehat{\Theta}^{-1}\,=\,-\, %
[\,-\widehat{V}\nabla_X\,+\,\widehat{\mathcal{L}}\,] %
\, \,\,, \label{ro}
\end{eqnarray}
where $\,\widehat{\Theta}\,$ %
is time reversal operator defined by
\begin{eqnarray}
\widehat{\Theta}\,\Delta_n(Y,K_n,\Sigma_n)\, %
=\, \Delta_n(-Y,-K_n,-\Sigma_n)\,\,\,\nonumber\\
\widehat{\Theta}\,\Delta\{Y,z_{\sigma}(k)\}\, %
=\, \Delta\{-Y,z_{-\sigma}(-k)\}\,
 \label{rr}
\end{eqnarray}
in the coordinate representation and by
\begin{eqnarray}
\widehat{\Theta}\,\Delta_n(p,K_n,\Sigma_n)\, %
=\, \Delta_n(-p,-K_n,-\Sigma_n)\,\,\,,\nonumber\\
\widehat{\Theta}\,\Delta\{p,z_{\sigma}(k)\}\, %
=\, \Delta\{-p,z_{-\sigma}(-k)\}\,
 \label{rp}
\end{eqnarray}
in the Wigner's one (in any time-reversed process, %
all momenta of particles and quanta are inverted %
while their creations are replaced by annihilations %
and vice versa). %
Hence, if $\,\Delta_n(t,X,Y,K_n,\Sigma_n)\,$ and %
$\,\Delta\{t,X,Y,z_{\sigma}(k)\}\,$ are some solutions %
of Eqs.\ref{ne} and \ref{fe}, respectively, %
then  $\,\widehat{\Theta}\, %
\Delta_n(t_0-t,X,Y,K_n,\Sigma_n)= %
\Delta_n(t_0-t,X,-Y,-K_n,-\Sigma_n)\,$ and %
$\,\widehat{\Theta}\, %
\Delta\{t_0-t,X,Y,z_{\sigma}(k)\}= %
\Delta\{t_0-t,X,-Y,z_{-\sigma}(-k)\}\,$ %
also are their solutions. %
As the consequence, the equilibrium stationary %
solution of evolution equations satisfies %
\begin{eqnarray}
\widehat{\Theta}\,\Delta_n^{eq}\,=\,\Delta_n^{eq}\, %
\label{eqs}
\end{eqnarray}

The properties (\ref{ho}) and (\ref{ro}) %
are characteristic principal properties of (quantum) %
Liouville operators. %
This is quite natural:\, %
relations (\ref{fmr}), (\ref{fmp}) and (\ref{fmo}) %
do show that the evolution %
operators\, $\,-\widehat{V}\nabla_X+\widehat{\Lambda}\,$\, %
and $\,-\widehat{V}\nabla_X+\widehat{\mathcal{L}}\,$\, %
are nothing but particular representations of the full %
system's Liouville operator, i.e.\, %
$\,[H,\dots]/i\hbar\,$.

This means that %
operators\, $\,-\widehat{V}\nabla_X+\widehat{\Lambda}\,$\, %
and $\,-\widehat{V}\nabla_X+\widehat{\mathcal{L}}\,$\, %
have the same spectrum as $\,[H,\dots]/i\hbar\,$ do. %
That is all their eigenvalues are purely imaginary, %
since spectrum of $\,[H,\dots]/i\hbar\,$ consists of %
$\,(E_S-E_{S^\prime})/i\hbar\,$\,, with\, $\,E_S\,$\, %
being eigenvalues of the full Hamiltonian:\, %
$\,H|S\rangle =E_S|S\rangle\,$. %
According to relation (\ref{fmo}), the corresponding %
eigenvectors of %
$\,-\widehat{V}\nabla_X+\widehat{\mathcal{L}}\,$\, %
and thus $\,-\widehat{V}\nabla_X+\widehat{\Lambda}\,$\, %
are determined by expression %
\begin{eqnarray}
\Delta_{S\,S^\prime}\{X,Y,z\}\,=\,%
\mathcal{F}^{-1}_{free}\{z\}\,\,\Tr_{\,ph}\, %
\exp{\left[\int %
z_-(k) \,a_k\,e^{ikr} \,d^3k\,\right]}\, %
\times\nonumber\\
\times \,\langle r|\,|S\rangle\langle S^\prime | %
\,|r^\prime\rangle \, %
\exp{\left[\int %
z_+(k) \,a_k^\dagger\, %
e^{-ikr^\prime} \,d^3k\,\right]}\, %
\, \, \label{ev}
\end{eqnarray}

Clearly, the functionals introduced by this %
expression form definite complete set %
of eigenfunctions of our evolution operators. %
Then the imaginary character of corresponding %
spectrum means that these operators can not %
have real eigenvalues or complex ones with nonzero  %
real parts. Next, we will consider most principal %
and important consequence of this circumstance, %
while Appendix A contains a its simple %
visual illustration.

\section{Joint characteristic function %
of total electron's path %
and instant system's state, and their cumulants}

Let initially the electron is located in some finite %
spatial region around the coordinate origin, %
so that, for example,
\begin{eqnarray}
\Delta_0^0(X,Y=0)\,=\,W_0(X)\,=\, %
(2\pi x_0^2)^{-3/2}\,\exp{(-X^2/2x_0^2)}\,\, \nonumber
\end{eqnarray}
in (\ref{icd}) and (\ref{icd1}), respectively, %
with some reasonable uncertainty $\,x_0\,$ %
(e.g. $\,x_0^2=\hbar^2/4mT\,$\, [1]). %
In the first case, the electron  %
is suddenly thrown in the phonon gas just %
at initial time moment $\,t=0\,$ and therefore %
at $\,t=0\,$ has no correlation with phonons. %
In the second case, at $\,t=0\,$ it already %
possesses all equilibrium correlations with %
phonons correspondingly to momentum and energy %
balance with them. %

This difference, however, must become unimportant %
after sufficiently large time, $\,t\gg\tau_0\,$, %
where $\,t\gg\tau_0\,$ is characteristic time %
of relaxation (thermalization) of electron's %
momentum probability distribution. %
Therefore at $\,t\gg\tau_0\,$, - %
or, formally, at $\,t\rightarrow\infty\,$, - %
we can concentrate %
on statistics of electron's random walk, %
i.e. on probability distribution, %
$\,W(t,X)\,$, of its %
coordinate $\,X=X(t)\,$. At that, since the %
latter unboundedly spreads in space, - %
so that $\,\langle X^2(t)\rangle %
/x_0^2,\rightarrow\infty\,$, - %
we can treat $\,X(t)\,$ as total %
electron's path accumulated  %
during time interval $\,(0,t)\,$. %

Then, as usually, it is convenient to investigate %
firstly characteristic function of the path distribution, %
by considering Fourier transforms
\[
\begin{array}{l}
\Delta_n(t,i\kappa,Y,K_n,\Sigma_n)\,=\, %
\int \exp{(i\kappa X)}\, %
\Delta_n(t,X,Y,K_n,\Sigma_n)\,d^3X\,\,\,,\\
\Delta\{t,i\kappa,Y,z\}\,=\, %
\int \exp{(i\kappa X)}\, %
\Delta\{t,X,Y,z\}\,d^3X\,\,\,,
\end{array}
\]
so that the evolution equations take form
\begin{eqnarray}
\dot{\Delta}_n \, =\,  %
i\kappa\,\widehat{V} \, \Delta_n\, %
+\,\sum_{k\,\in\,K_n}\,\widehat{L}_{k\,\sigma_k}\, %
\Delta_n \,+\,\nonumber\\ %
+\,\sum_{k\,\in\,K_n}\,\widehat{B}_{k\,\sigma_k}\, %
\Delta_{n-1}\,+\, %
\sum_{\sigma\in \{+,-\}}\,\int %
\widehat{A}_{q\,\sigma}\, %
\Delta_{n+1}\, d^3q\, \,\, \label{nek}
\end{eqnarray}
(where, of course,\, $\,K_{n-1}=K_n\ominus k\,$, %
$\,\Sigma_{n-1}=\Sigma_n\ominus \sigma_k\,$,%
$\,K_{n+1}=K_n\oplus q\,$ and, %
$\,\Sigma_{n+1}=\Sigma_n\oplus \sigma\,$), %
- or shortly\,
\begin{eqnarray}
\dot{\Delta}  \,=\, %
[\,i\kappa\,\widehat{V}\, +\, \widehat{\Lambda}\,] %
\,\Delta\,\,\,,\label{vek}
\end{eqnarray}
- and %
\begin{eqnarray}
\dot{\Delta} \, =\,  %
i\kappa\,\widehat{V}\, \Delta\, %
+\,\widehat{\mathcal{L}}\,\Delta\,\,\,,\label{fek}
\end{eqnarray}
with operators defined in (\ref{fe})-(\ref{fops}). %

According to Eq.\ref{fmr}, %
\begin{eqnarray}
\Delta\{t,i\kappa,Y,z\}\, %
=\, \frac {\mathcal{F}\{t,i\kappa,Y,z\}} %
{\mathcal{F}_{free}\{z\}}\,\,\,,\label{fmk}
\end{eqnarray}
where
\begin{eqnarray}
\mathcal{F}\{t,i\kappa,Y,z\}\,\equiv\, %
\int \exp{(i\kappa X)}\, %
\mathcal{F}\{t,X,Y,z\}\,d^3X\,=\,\label{ffk}\\ %
=\,\left\langle\, %
\exp{\left[\,i\kappa\,X(t)\,+\, %
i\xi V(t)\, + \sum_\sigma \int %
z_\sigma(k) A^\sigma_k(t) \,d^3k\,\right]} %
\,\right\rangle\,\,\,, \nonumber
\end{eqnarray}
with\, $\,V(t)=p(t)/m\,$\, being electron's %
velocity, and
\[
\begin{array}{l}
\xi\,\equiv\, mY/\hbar\,
\end{array}
\]
Thus,\, $\,F\{t,i\kappa,Y,z\}\,$\, %
in fact is full characteristic function %
of our system (in the sense of the %
probability theory), %
while $\,\Delta\{t,i\kappa,Y,z\}\,$\, %
differs from it only by exclusion of %
undisturbed part of self-correlations %
of phonon modes. %
At the same time, since
\[
\begin{array}{l}
X(t)\,=\,X(0)\,+\int_0^t V(\tau)\,d\tau\,
\end{array}
\]
and\, $\,|X(0)|\lesssim x_0\,$, function %
$\,\Delta\{t,i\kappa,Y,z\}\,$\, serves as %
characteristic functional of electron's %
velocity on the interval $\,(0,t)\,$.

As usually, it is reasonable to %
consider this function in terms of various %
cumulants and their generating functional: %
\begin{eqnarray}
\mathcal{C}\{t,i\kappa,\xi,z\}\,\equiv\,\ln\, %
\Delta\{t,i\kappa,Y,z\}\, =\, \label{cum}\\ %
=\, %
\sum_{j+l+n\,>\,0}\, \frac {(i\kappa)^j %
(i\xi)^l}{j!l!n!} %
\sum_{\sigma_1\dots \sigma_n}\, \int_{k_1} %
z_{\sigma_1}(k_1) \dots \int_{k_n} %
z_{\sigma_n}(k_n)\,\, %
C_{jln}(t,K_n,\Sigma_n)\,\,\,,\nonumber\\
C_{jln}(t,K_n,\Sigma_n)\,=\, %
\left\langle\left\langle\,X^j(t)\,V^l(t)\, %
\prod_{s\,=\,1}^n\, A^{\sigma_s}_{k_s}(t)\ %
\right\rangle\right\rangle %
\,\,\,, \nonumber
\end{eqnarray}
where $\,\int_k\dots =\int\dots d^3k\,$, %
$\,K_n=\{k_1\dots k_n\}\,$, %
$\,\Sigma_n=\{\sigma_1\dots \sigma_n\}\,$,\, %
the double angle brackets denote %
$\,(j+l+n)\,$-order cumulant, %
i.e. purely irreducible %
part of  $\,(j+l+n)\,$-order %
mutual correlation, of %
the enveloped multipliers,\, %
and equity of some of indices $\,j,l,n\,$ %
to zero means absence of corresponding %
multipliers (only in special case, when %
$\,j=l=0,\, n=2\,$,\, $\,C_{002}\,$ %
is difference between full cumulant %
and its ``seed'' value determined by\, %
$\,\ln\,\mathcal{F}_{free}\,$). %
In particular,
\begin{eqnarray}
\mathcal{C}\{t,i\kappa,\xi,0\}\,=\,\ln\, %
\Delta_0(t,i\kappa,Y)\,=\,  \label{cum0}\\ %
=\, %
\sum_{j+l\,>\,0}\, \frac {(i\kappa)^j %
(i\xi)^l}{j!l!} \,  %
\left\langle\left\langle\,X^j(t)\,V^l(t)\, %
\right\rangle\right\rangle %
\, \nonumber
\end{eqnarray}
describes most interesting for us %
full joint statistics of %
electron's velocity and total path %
in themselves.

Notice that in terms of cumulants Eq.\ref{fek} %
becomes nonlinear:
\begin{eqnarray}
\dot{\mathcal{C}} \, =\,  %
\left[\,i\kappa\,-\, %
\sum_\sigma \int d^3k\,\,\, %
z_\sigma(k)\, %
i\sigma k\,\frac %
{\delta \,\mathcal{C}}{\delta z_\sigma(k)}\, %
\right]\,%
\widehat{V}\,\mathcal{C}\,+\, %
\label{cek}\\
+\, %
\sum_\sigma \int d^3k\,\,\, z_\sigma(k)\, %
\widehat{L}_{k\,\sigma}\,\frac %
{\delta \,\mathcal{C}}{\delta z_\sigma(k)}\, %
+\, \nonumber\\
+\, \sum_\sigma \int d^3k\,\,\, z_\sigma(k)\, %
\widehat{B}_{k\,\sigma}\, +\, %
\sum_\sigma \int d^3k\,\,\, %
\widehat{A}_{k\,\sigma}\,\frac %
{\delta \,\mathcal{C}}{\delta z_\sigma(k)}\, %
\,\,, \nonumber
\end{eqnarray}
with quadratic nonliearity.

\section{Long-time asymptotic of the cumulants %
and electron's diffudivity/mobility low-frequency %
fluctuations}

Now, let us consider long-range time behavior %
of the electron's path and related %
cumulants at $\,t/\tau_0 %
\rightarrow\infty\,$. %

Our sole assumption will be that electron's %
interaction with the phonon thermostat ensures %
fast enough, - i.e. integrable, - relaxation %
of the electron's velocity correlation function\, %
$\,\langle\langle\, V(t)V(t^\prime)\, %
\rangle\rangle\,$\, and thus diffusive character %
of the random walk, in the sense that %
\begin{eqnarray}
C_{200}(t)\,=\, %
\langle\langle\, X^2(t)\, \rangle\rangle\, %
\rightarrow\,\,2Dt\,\, +\,\texttt{const} \,\, %
\label{dif}
\end{eqnarray}
Undoubtedly, this is the case under suitable %
choice of electron-phonon couplings $\,|c_k|^2\,$ %
and phonon frequencies $\,\omega_k\,$\,.  %
In particular, without essential loss of generality %
we may take  %
\begin{eqnarray}
c_k\,=\,c(|k|)\,\,\,, \,\,\,\,\,\,
\omega_k\,=\,\omega(|k|)\,\,\,,
\label{ss}
\end{eqnarray}
with $\,c(|k|)=c^*(|k|) \ge 0\,$ %
(phases of $\,c_k\,$'s anyway are %
of no importance). This guarantees spherical symmetry %
of the random walk. %

A suitable choice can be recognized from the simplest %
approximation of exact equations, i.e. the ``kinetic %
equation'', - see Eq.58 in Sec.5.4 in [1]. %
In the representation under %
consideration it can be written as
\begin{eqnarray}
\dot{\Delta_0}(t,i\kappa,Y)\,=\, %
[\,i\kappa \widehat{V}\,+\, %
\widehat{K}\,]\,\Delta_0(t,i\kappa,Y) %
\,\,\,,\label{ke}\\ %
\widehat{K}\,\dots\,\equiv\, %
-\sum_{\sigma}\int_k \widehat{A}_{k\sigma}\, %
[\widehat{L}_{k\sigma}-\epsilon]^{-1}\, %
\widehat{B}_{k\sigma}\,\dots\,\,\,, \nonumber
\end{eqnarray}
with\, $\,\epsilon =+0\,$\,\footnote{\, %

Notice that $\,\widehat{L}_{k\sigma}\,$ %
is imaginary-valued operator. Addition of %
infinitely small real %
positive\, $\,\epsilon =+0\,$ to it is necessary %
for artificial breaking of the time reversal %
symmetry. Otherwise, at $\,\epsilon =0\,$, %
as one can easy verify, the integral %
in (\ref{ke}), i.e. $\,\widehat{K}\,$, %
would turn to zero (see also Appendix A %
below).}\,. %

Provided the mean square of electron's path %
obeys the diffusive law (\ref{dif}), %
let us discuss possible long-term asymptotic of %
higher-order path cumulants and other cumulants. %

If we were basing on habitual intuition, %
in turn based on standard kinetic theory, %
we would predict that all the cumulants %
from (\ref{cum}) and (\ref{cum0}) tend %
to finite limits, except self-cumulants %
of the electron's path:
\begin{eqnarray}
C_{sln}(t)\,\rightarrow\,\texttt{const}\,\,\,\, %
\,\,\,\, %
\texttt{if}\,\,\,\,\,l+n>0\,\,\,,\label{ass0}\\
C_{s00}(t)\,=\,
\left\langle\left\langle\,X^s(t)\, %
\right\rangle\right\rangle %
\rightarrow\,sD_s\,t\,+\,\texttt{const} %
\,\,\,, \label{ass}
\end{eqnarray}
where
\begin{eqnarray}
D_s\,=\frac 1s %
\,\frac d{dt}\,%
\left\langle\left\langle\,X^s(t)\, %
\right\rangle\right\rangle\,=\, %
\left\langle\left\langle\,V(t)X^{s-1}(t)\, %
\right\rangle\right\rangle\, %
\rightarrow\,\texttt{const} %
\,\, \nonumber
\end{eqnarray}
and\, $\,D_2=D\,$\, %
(of course, $\,D_s =0\,$\, at odd $\,s\,$\,  %
because of the spherical symmetry). %
Such behavior means that, %
at any\, $\,n>0\,$\, and any  %
$\,\eta_j(t)\,$ being either $\,V(t)\,$ %
or some of $\,A_k^\sigma(t)\,$\,, %
\begin{eqnarray}
\left\langle\left\langle\, %
\eta_1(t)\dots \eta_s(t) \, %
V(t_1)\dots V(t_n)\, %
\right\rangle\right\rangle %
\,\rightarrow\,0 \,\,\, \label{ass1}\\ %
\texttt{when some of}\,\,\,\,\,\,\, %
t-t_j \rightarrow\infty\,\,\,, \nonumber
\end{eqnarray}
where the zero limit is achieved %
in so fast way, that these cumulants are %
integrable over all $\,t_j\,$ simultaneously. %
Orally, all irreducible correlations between %
present state of the system, - %
characterized by electron's velocity and %
phonon amplitudes, - and past values of %
velocity disappear with time. %
Then we can write
\begin{eqnarray}
\Delta\{t,i\kappa,Y,z\}\,=\, %
e^{\,\lambda(i\kappa)\,t}\,\,%
\Delta^\prime \{t,i\kappa,Y,z\}\, %
\rightarrow\, %
e^{\,\lambda(i\kappa)\,t}\,\,%
\Delta^\prime \{\infty,i\kappa,Y,z\}\,\,\,, %
\label{ass2} %
\end{eqnarray}
with
\begin{eqnarray}
\lambda(i\kappa)\,=\, \lim_{t\rightarrow\infty} %
\,\frac {\mathcal{C}\{t,i\kappa,0,0\}}t %
\,=\,\sum_{s=1}^\infty\, %
\frac {D_{2s}}{(2s-1)!}\, (-\kappa^2)^s\,%
<\,0\,\,\, \nonumber
\end{eqnarray}
Correspondingly, the path probability distribution %
is asymptotically Gaussian:
\begin{eqnarray}
W(t,X)\,\rightarrow\,(4\pi Dt)^{-3/2} %
\exp{(-X^2/4Dt)}\,\,\, \label{ga}
\end{eqnarray}
Just such type of %
asymptotical behavior follows from %
the kinetic equation.
At that, $\,\lambda(i\kappa)\,$ is real %
negative quantity determined by the eigenvalue problem %
\[
\begin{array}{l}
\lambda(i\kappa)\,f(i\kappa,Y)\,=\, %
[\,i\kappa \widehat{V}\,+\,\widehat{K} %
\,]\, f(i\kappa,Y)\,\,\,
\end{array}
\]

\,\,\,

But, unfortunately, %
from the viewpoint of exact equations, %
(\ref{nek}) or (\ref{vek}) or (\ref{fek}), %
so nice behavior is impossible! %

\,\,\,

Indeed, if the assumed asymptotic %
(\ref{ass0}),(\ref{ass}),(\ref{ass2}) was true, then, %
evidently, the Eq.\ref{fek} would result in equality %
\begin{eqnarray}
[\,i\kappa\widehat{V}\, +\, %
\widehat{\mathcal{L}} %
\,]\, %
\Delta^\prime \{\infty,i\kappa,Y,z\}\, %
=\, \lambda(i\kappa)\, %
\Delta^\prime \{\infty,i\kappa,Y,z\}\,\,\,, %
\label{aslim} %
\end{eqnarray}
thus stating that operator\, %
$\,i\kappa\widehat{V}+ \widehat{\mathcal{L}}\,$\, %
has a real eigenvalue. %
Then this would imply, - as it is easy to see, - %
that operator %
\,$\,-\widehat{V}\nabla_X+\widehat{\mathcal{L}}\,$\, %
(and, equivalently,\, %
\,$\,-\widehat{V}\nabla_X+\widehat{\Lambda}\,$) %
has the same real eigenvalue, %
\begin{eqnarray}
\lambda(i\kappa)\,f\{X,Y,z\}\,=\, %
[\,-\widehat{V}\nabla_X\,+\,\widehat{\mathcal{L}}\,]\, %
f\{X,Y,z\}\,\,\,, \label{asev} %
\end{eqnarray}
with corresponding eigenvector expressed by
\begin{eqnarray}
f\{X,Y,z\}\,=\, e^{-i\kappa X}\,%
\Delta^\prime \{\infty,i\kappa,Y,z\}\,\,\, %
\label{asef} %
\end{eqnarray}
However, this is impossible, since, - %
as we have shown above, - the whole spectrum  %
of the exact evolution operator is purely imaginary.
It is useful to add %
that this reasoning stays valid even if the %
constants in (\ref{dif}) and (\ref{ass}) are replaced %
by arbitrary sub-linear time functions.

\,\,\,

Hence, the assumptions %
(\ref{ass0}) and (\ref{ass}) (at\, $\,s>2\,$), %
and thus (\ref{ass2}) and the hypothetical asymptotic (\ref{ga}), %
in spite of their seeming plausibility, all are qualitatively %
wrong.

One  more proof of incompatibility %
of these assumptions with exact %
evolution equations comes from  %
the property (\ref{ro}). %
The latter implies, firstly, %
that if (\ref{asef}) is eigenfunction %
of the operator\, %
$\,-\widehat{V}\nabla_X+\widehat{\mathcal{L}}\,$, %
with eigenvalue\, $\,\lambda(i\kappa)\,$, %
then this operator has also eigenvalue\, %
$\,-\lambda(i\kappa)\,$, corresponding to %
eigenfunction %
$\,e^{-i\kappa X}\, %
\widehat{\Theta}\, %
\Delta^\prime \{\infty,i\kappa,Y,z\}\,$. %
Therefore, secondly, %
starting at $\,t=0\,$ from initial conditions %
(e.g. (\ref{icd}) or (\ref{icd1})) what satisfy %
time-reversal symmetry, like (\ref{eqs}), %
we must come to asymptotic which, %
in contrast to (\ref{ass2}), %
with equal rights includes two %
exponentials,\, $\,\exp{[\lambda(i\kappa)t]}\,$\, %
and\, $\,\exp{[-\lambda(i\kappa)t]}\,$\,, and %
thus is certainly physically senseless! %

\,\,\,

Consequently, we have to conclude that %
in reality some of the ccumulants in (\ref{ass1}) %
are not fast enough decaying, i.e. integral of some %
of them over all $\,t_j\,$ from $\,0\,$ to $\,t\,$ %
are unboundedly growing with time $\,t\,$. %
The natural first candidate for such role %
is\,\footnote{\,

The cumulant %
$\,\left\langle\left\langle\, %
V(t) V(t_1)\, \right\rangle\right\rangle\,$, %
can not play such role as far as the mean square %
diffusive law is fulfilled, either in the %
form (\ref{dif}) or\, %
$\,\, \left\langle\left\langle %
X^2(t) \right\rangle\right\rangle /t\, %
\rightarrow\,2D\,=\,$const\,. %
}\,
\[
\begin{array}{l}
\left\langle\left\langle\, %
V(t) \, V(t_1)V(t_2) V(t_3)\, %
\right\rangle\right\rangle\,\,\,, %
\end{array}
\]
that is
\begin{eqnarray}
\left\langle\left\langle\, %
V(t) \,X^3(t)\, %
\right\rangle\right\rangle\, %
\rightarrow\,\infty \,\,\, \label{ass11}
\end{eqnarray}
This implies that
\begin{eqnarray}
\frac {\left\langle\left\langle\, %
X^4(t) \,\right\rangle\right\rangle}t\, \,=\, %
\frac { 4\int_0^t \left\langle\left\langle\, %
V(t)X^3(t) \,\right\rangle\right\rangle %
\,dt}{t}\, \, %
\sim \,\, %
\left\langle\left\langle\, %
V(t) \,X^3(t)\, %
\right\rangle\right\rangle\, %
\rightarrow\,\infty \,\,\, \label{ass12}
\end{eqnarray}
Thus, the \,{\bf fourth-order cumulant of %
electron's path grows with time %
obeying a super-linear law}.
This conclusion  is main final %
result of the present paper.

We see that long-range behavior of (even equilibrium) %
electron's random walk can not be adequately %
described with the help of single parameter\, %
$\,D\,$\,, which now determines mean square %
value of path only, but not its %
higher-order statistical moments.  %

\,\,\,

Further consequencies of this  result, concerning %
the 1/f\,-noise, do follow already known %
general relations \cite{pr1,bk1,bk3,i1,i2}. %
Namely, the asymptotic (\ref{ass12}) shows that %
the electron's path\, $\,X(t)\,$\, behaves %
as if electron's diffusivity was a random variable, %
$\,\widetilde{D}(t)\,$, with mean value $\,D\,$ %
and effective correlation function %
\begin{eqnarray}
\left\langle\left\langle\, %
\widetilde{D}(t)\widetilde{D}(0)\, %
\right\rangle\right\rangle \,=\, %
\frac 1{24}\, \frac {d^2}{dt^2} \, %
\left\langle\left\langle\,X^4(t)\, %
\right\rangle\right\rangle\,=\, %
\frac 12\, %
\left\langle\left\langle\,V(t)X^2(t)V(0)\, %
\right\rangle\right\rangle\,
\label{dcf}
\end{eqnarray}
From (\ref{ass11})-(\ref{ass12}) one finds %
that this ``diffusivity correlation function'' %
is slowly decaying (definitely non-integrable) %
time function\,\footnote{\,

Formulae like (\ref{dcf}) appeared in %
\cite{pr1,bk1,bk3}. %
They presume that observation of (equilibrium) %
random path starts from nearly equilibrium %
initial condition like our (\ref{icd1}). %

One of possible forms %
of the diffusivity correlation function %
was predicted and considered in  \cite{pr1,bk1,bk3,i2}. %
In principle, generally speaking, %
this  function, being introduced %
by (\ref{dcf}), can be %
even not decaying but instead growing with time. %
Such possibility was demonstrated and explained %
from physical point of view in %
\cite{i1,i2,p1,tmf,lpro}. %
}\,. %

Analogous consideration of our equations %
in presence of external force (as %
was mentioned above) %
or, alternatively, application to our results %
of the generalized fluctuation-dissipation relations %
shows that the diffusivity fluctuations always  %
convert themselves into similar fluctuations %
of (low-field) electron's mobility, with the same %
correlation function (\ref{dcf})\,\,\footnote{\, %

See \cite{bk2,bk3,i1,i2,jstat,last} and, for %
the fluctuation-dissipation relations, %
references mentioned in \cite{bk3,i1,i2,i3} %
and in preprints [Yu.E.Kuzovlev, %
arXiv: 1106.0589-,, 1108.1740\,]. %
}\,.

Obviously, the corresponding spectrum %
of diffusivity/mobility fluctuations %
is of 1/f type\,\footnote{\, %

For details and examples see e.g. %
\cite{bk1,bk3,i1,i2,p1,tmf,jstat}. %
}\,\, - %
or, more generally, of 1/f$^\gamma\,$ type, - %
showing unbounded growth at %
$\,f\tau_0 \rightarrow 0\,$. %
A concrete form of this spectrum produced %
by our exact equations will be investigated %
separetely.

\section{Discussion}

As it was realized already in %
\cite{pr1,bk1,bk2,bk3,i1,pr2}, %
low-frequency 1/f\,-type fluctuations %
in diffusivity and mobility of charge %
carriers, - as well as other %
molecular-size (quasi-) particles, - %
originates from indifference of many-particle %
systems to a number and relative %
frequency of collisions (elementary %
acts of interactions, scattering, etc.) %
of any concrete particle. In other %
words, indifference of many-particle systems %
to rates of participation of a concrete particle %
in various irreversible and random processes, %
first of all, in the particle's %
thermalization (e.g. momentum and energy %
relaxation), diffusion and drift under external %
forces.

Let, for example, during last 1 second %
some air particle or charge carrier has undergone %
two times greater (or lesser) number %
of collisions than it must have ``on average''. %
Such incident will make no influence upon %
next particle's motion, %
since any detail of its motion becomes %
forgotten by the system after %
characteristic time $\,\tau_0\,$ %
much shorter than 1 second. %
In other words, even strong fluctuation of %
number of collisions, - and hence of rate of %
particle's diffusion, - does not cause a %
``back reaction'' of the system %
(which would enforce this particle %
to make ``opposite fluctuation'' in %
number of collisions during next 1 second %
and ``compensate'' the old incident).

This reasoning shows principal difference of fluctuations %
in relative frequency of collisions (and  %
connected physical quantities) %
from, say, particle's energy fluctuations %
(which cause back reaction suppressing %
energy deviations from its mean value %
during time $\,\sim\tau_0\,$).

The 1 second in above reasoning can be %
replaced by arbitrary time interval %
greater than $\,\tau_0\,$, and %
instead of 1 second we can take %
10 seconds, 100 seconds, %
and so on\,\,\footnote{\, %

Strictly speaking, this interval %
can be bounded above by some %
characteristic ``non-ergodicity time'', %
which, however, is very large for real, %
even closed, many-particle systems %
\cite{tmf}) and infinitely large for %
open or infinitely many-particle %
systems.}\,.\, %
At that, because the system constantly %
forgets history of the particle's collisions, %
it, figuratively speaking, can not %
distinguish what part of collisions %
is ``average'' and what is ``deviation %
from average''. Hence, it allows the latter %
to be as fast growing with time as the former. %
This characteristic law,\, %
$\,N(t)-\langle N(t)\rangle \propto %
\langle N(t)\rangle \propto t\,$ %
(with $\,N(t)\,$ standing number of collisions) %
just means that spectrum of fluctuations %
of instant relative frequency of collisions %
(and thus of rate of diffusion, etc.) %
at low frequencies $\,f\ll 1/\tau_0\,$ %
has nearly 1/f\, form (differing from it %
by some logarithmic factors). %

Additional explanations of so simple origin %
of 1/f\,-noise can be found in %
\cite{july,i2,p1,tmf,jstat,last,i3,kmg,lpro}. %

Quite obvious conclusion from the aforesaid %
is that theoreticians should honestly %
investigate actual Hamiltonian dynamics %
of many-particle systems and their dynamical %
chaos, - basing on exact %
equations of statistical mechanics, - %
instead of thoughtless demonstration %
of Bernoulli's ``art of conjectures''. %

In essence, the same was claimed %
by N.\,Krylov \cite{kr} as far ago %
as in 1950. One of important results of %
\cite{kr} is statement that even in ``good'' %
Hamiltonian systems (possessing excellent %
mixing properties) relative frequencies %
of a given sort of events %
on different phase trajectories, - %
i.e. in different experiments, - %
generally differ one from another %
and thus from ensemble average, %
regardless of duration of experiments. %
Therefore, - as N.\,Krylov underlined, - %
it is impossible to describe actual %
randomness of physical systems in terms %
of any a priori introduced ``probabilities %
of events'' (e.g. collisions). %
In other words, ``probabilities %
of events'' have no certain values. %

Hence, all probability-theoretical models %
of noise inevitably lose fundamental %
(in the above explained sense) 1/f\,-noise. %
That is why we have to investigate %
dynamical models without any a priori %
truncation of their exact equations %
(for any truncation, - like one leading %
to the ``kinetic equation'' (\ref{ke}), - %
acts as artificial introduction of some %
a priori probabilities).

One more important Krylov's result is %
that in statistical mechanics %
statistical correlations between %
particles or events do not necessarily %
say about some real ``physical'' %
(cause-and-consequence) correlations or %
connections between them. %
Our above treatment of collisions' %
number fluctuations well illustrates this %
fact. %
Thus it would be wrong to interpret %
the 1/f\,-noise under discussion %
as manifestation of some physical %
long-living correlations %
(e.g. between charge carriers) or %
some ``slow relaxation''. %
There are only purely statistical %
correlations, and they manifest %
only so rich randomness which can not %
be adequately imitated by traditional %
probability-theoretical %
(kinetic and stochastic) models. %

\,\,\,

The situation with 1/f\,-noise in %
semiconductors \cite{hkv,fnh} %
serves as remarkable example %
(see \cite{bk3,i2,last}). %
On one hand, many experiments %
give strong evidences that main source of this noise %
is (independent) 1/f\,-type fluctuations %
of mobilities of charge carriers resulting %
from their scattering by lattice vibrations, %
i.e. phonon gas \cite{hkv,fnh}. %
On the other hand, the traditional %
way of thinking enforces to switch attention %
from the electron-phonon interaction and %
scattering itself to the phonon subsystem %
in itself and suggest 1/f\, fluctuations of %
occupancies of phonon modes %
(see e.g. \cite{fnh} and references therein). %
Even in spite of clear contradiction between %
such idea and kinetic theory which establishes %
quite definite (finite) relaxation times of %
phonon modes!\,\,\footnote{\,  %

Thus authors of such idea are even more %
``revolutionary'' than I am! I do not suggest that %
results of kinetic theory are invalid, stating only %
that they are %
incomplete, and  that, to supplement them properly, %
one should return %
from habitual probability-theoretical scheme
\[
\begin{array}{l}
Probabilities\,\,\,\Rightarrow\,\,\, %
\stackrel{\texttt{Stochastic}}{\texttt{model}}\, %
\,\,\,\Rightarrow\,\,\, Noise\, %
\end{array}
\]
to canonical statistical-mechanical %
scheme
\[
\begin{array}{l}
\stackrel{\texttt{Hamiltonian}}{\texttt{model}}\, %
\,\,\,\Rightarrow\,\,\, Noise\, %
\,\,\,\Rightarrow\,\,\, Probabilities\, %
\end{array}
\] %
}\,\,.\, %
In fact, however, low-frequency 1/f\,-type %
fluctuations observed in phonon systems themselves %
(e.g. fluctuations of internal friction in quartz %
crystals or intensity of light scattering by them) %
come from fluctuations in relaxation %
rates of phonon modes there (see \cite{i3} and %
references therein), - %
in accordance with the above reasonings, - %
but in no way from fluctuations of %
phonon numbers. %

In our Hamiltonian model, %
considered in this paper, %
perturbation of any particular phonon mode by the %
electron vanishes under the thermodynamic %
limit. %
In other words, any finite sequence of %
$\,\sim t/\tau_0\,$ %
electron's scattering events during finite %
time $\,t\,$ involves only %
vanishingly small portion of total number %
of phonon modes. Therefore,  %
all their occupancies do not fluctuate at all,  %
staying mere random constants. %
At that, since they are independent one on another, %
all effects of their randomness are suppressed %
by the infiniteness of their density in %
$\,k\,$-space (any finite volume there includes %
infinitely many modes). %
Hence, equations of our model describe just what is %
most interesting for us and for applications, %
i.e. 1/f\,-noise resulting from uncertainty of rate of %
electron's scattering by phonons. %

Of course, real samples under  measurements are %
finite and rather small, %
but one should remember about very many degrees of %
freedom in their surroundings which inevitably %
somehow interact with any sample. %

\section{Conclusion}

To resume, we have shown that %
if an interaction of electron (quantum particle) %
with ideal phono gas (harmonic boson thermostat) %
enforces the electron to move like Brownian %
particle, - so that mean square of its path %
obeys the diffusion law,\, %
$\,\langle X^2(t)\rangle\,\propto \,t\,$\,, - %
then such interaction certainly supplies %
this motion with 1/f\, (or ``flicker'') low-frequency %
fluctuations in electron's diffusivity and %
thus mobility.

In essence, this our result was deduced %
from most general properties of exact %
statistical-mechanical equations of %
the electron-phonon system only. %
This fact prompts that our result %
can be extended to a wide class of %
different systems, in the form of some %
general theorem (which surely will include %
the molecular random walk in fluids considered %
in \cite{i1,tmf,jstat,lpro}). %
This seems very attractive task.
But not less intriguing task is concretization %
of our present result up to exact description %
of shape (frequency dependence) of the 1/f\,-type %
spectrum, along with its adequate enough %
quantitative estimates. I hope to contribute %
to such work.


\appendix

\section{Imaginary spectrum of the evolution %
operator and falsity  of the kinetic equation}

Notice that properties %
(\ref{ho}) and (\ref{ro}) remain formally valid under %
truncations of the evolution equations. Therefore %
character of spectrum of the evolution operator %
remains unchanged. And, in order to feel it, %
we can for simplicity make truncation already %
at $\,n=2\,$, by formal putting $\,\Delta_2=0\,$, %
as was done in [1] when deriving the ``kinetic equation'', %
Eq.58 in [1]. %
Besides, for more simplicity, we will confine %
ourselves by spatially homogeneous eigenfunctions. %

Thus, let us consider the eigenvalue problem %
\begin{eqnarray}
\lambda\,\Delta_0(Y)\,=\, %
- \,\frac 1{i\hbar}  %
\sum_{\sigma\in \{+,-\}}\, \sigma %
\int c_k^{\,\sigma}\, %
(\,e^{\,ikY/2}-e^{-ikY/2})\,\,  %
\Delta_{1}(Y,k,\sigma)\, \,\frac {d^3k}{(2\pi)^3}\,
\,\,,\nonumber\\
\lambda\,\Delta_1(Y,k,\sigma)\,=\, %
\frac 1{i\hbar}\, %
\sigma\, c_k^{-\sigma}\, %
[\,N_k\,e^{\,ikY/2}- (N_k +1)\,e^{-ikY/2}\,]\, %
\Delta_{0}(Y)\,+\,\nonumber\\
+\, i  \sigma\, %
(\omega_k\,-\,k\widehat{V})\, \Delta_1(Y,k,\sigma)\, %
\,\,, \label{a1}
\end{eqnarray}
and assume that it has solution %
with a real non-zero $\,\lambda\,$. %
From viewpoint of the kinetic %
equation this seems undoubted, because %
kinetic equation arose in %
the same approximation, and spectrum of %
its $\,X\,$-independent eigenfunctions (spectrum of %
momentum and energy relaxation) certainly is real.

Excluding $\,\Delta_1\,$ %
from this pair of equations and going to the Wigner %
representation, we come to %
\begin{eqnarray}
\lambda\,\Delta_0(p)\,=\, \label{a2}\\ %
\frac {2\pi}{\hbar^2}  \int |c_k|^2\, %
\{\, %
\delta_{\lambda}(\omega_k- kV +\frac {\hbar k^2}{2m}) %
\,\times \nonumber\\ %
\times \, %
[\,N_k\,\Delta_0(p-\hbar k) %
- \,(N_k+1)\, \Delta_0(p)]\,+\,\nonumber\\ %
\,+\,\, %
\delta_{\lambda}(\omega_k- kV -\frac {\hbar k^2}{2m}) %
\,\times\nonumber\\
\times\, [(N_k+1)\,\Delta_0(p+\hbar k) %
\,-\, N_k\, \Delta_0(p)\,] %
\,\} %
\, \frac {d^3k}{(2\pi)^3}\, %
=\,\, \nonumber\\
=\,\int [\,V_{\lambda}(p|p^\prime)\,\Delta_0(p^\prime)\, %
-\, V_{\lambda}(p^\prime|p)\,\Delta_0(p)\,]\, %
d^3p\prime \,\,\,, \nonumber
\end{eqnarray}
where
\begin{eqnarray}
\delta_{\lambda}(x)\,\equiv\, %
\frac 1\pi \, %
\frac {\lambda}{\lambda^2+x^2}\,\,\,,\label{a3}\\
V_{\lambda}(p|p^\prime)\,\equiv\, %
\frac {2\pi}{\hbar^2} \int |c_k|^2\, %
\delta(p-p\prime -\hbar k)\, [\, N_k\, %
\delta_{\lambda}(\omega_k + \hbar k^2/2m -kV)\, %
+\, \nonumber\\
+\,(N_k+1)\, \delta_{\lambda}(\omega_k - \hbar k^2/2m + kV)\, %
]\,\frac {d^3k}{(2\pi)^3}\,
\,\label{a4}
\end{eqnarray}

At $\,\lambda \rightarrow\,+0\,$\, %
operator on the right in Eq.\ref{a2} %
turns to the kinetic operator from the Eq.58 %
in [1], that is operator $\,\widehat{K}\,$ %
from the same Eq.\ref{ke}) above, %
whose spectrum is real non-positive. %
At finite $\,\lambda >0\,$\,, evidently,  %
the operator in Eq.\ref{a2} has the same %
structure as general kinetic operators, %
therefore it also is non-positively defined. %
Hence, Eq.\ref{a2} can not have non-trivial %
solution with $\,\lambda >0\,$. %
If $\,\lambda <0\,$ then signs of both sides of %
$\,\lambda <0\,$ do change simultaneously and %
thus remain opposite. Therefore %
$\,\lambda <0\,$ also can not give the solution. %
Clearly, addition to $\,\lambda \,$ %
of imaginary component does not change %
the situation. %
Consequently, all non-zero eigenvalues %
in the Eqs.\ref{a1} are purely imaginary.

This well illustrates that the kinetic equation, %
in spite of its usefulness, %
is in principal contrast with its parent %
evolution equations.


%

\,\,\,

---------------------------

\,\,\,

\end{document}